# A Computer-Aided Diagnosis System Using Artificial Intelligence for Hip Fractures -Multi-Institutional Joint Development Research-

Yoichi Sato M.D.[1,6], Yasuhiko Takegami M.D. Ph.D.[2], Takamune Asamoto M.D.[3,6], Yutaro Ono MD[4,6], Hidetoshi Tsugeno[4,6], Ryosuke Goto[5], Akira Kitamura M.S. in IT[5], Seiwa Honda B.L.[6]
1)Department of Orthopedics Surgery, Gamagori City Hospital, Japan
2) Department of Orthopaedic Surgery, Nagoya University Graduate School of Medicine, Nagoya, Japan.
3)Department of Orthopedics Surgery, Tsushima City Hospital, Japan
4)Department of Orthopedics Surgery, Nagoya Daini Red Cross Hospital, Japan
5)Search Space CO.Ltd., Japan
6) Nonprofit Organization (NPO) Nagoya Orthopedic Regional Healthcare Support Center, AI Research Division, Nagoya, Japan

## Abstract

[Objective]
To develop a Computer-aided diagnosis (CAD) system for plane frontal hip X-rays with a deep learning model trained on a large dataset collected at multiple centers.
[Materials and Methods].
We included 5295 cases with neck fracture or trochanteric fracture who were diagnosed and treated by orthopedic surgeons using plane X-rays or computed tomography (CT) or magnetic resonance imaging (MRI) who visited each institution between April 2009 and March 2019 were enrolled. Cases in which both hips were not included in the photographing range, femoral shaft fractures, and periprosthetic fractures were excluded, and 5242 plane frontal pelvic X-rays obtained from 4,851 cases were used for machine learning. These images were divided into 5242 images including the fracture side and 5242 images without the fracture side, and a total of 10484 images were used for machine learning. A deep convolutional neural network approach was used for machine learning. Pytorch 1.3 and Fast.ai 1.0 were used as frameworks, and EfficientNet-B4, which is pre-trained ImageNet model, was used. In the final evaluation, accuracy, sensitivity, specificity, F-value and area under the curve (AUC) were evaluated. Gradient-weighted class activation mapping (Grad-CAM) was used to conceptualize the diagnostic basis of the CAD system.
[Results]
The diagnostic accuracy of the learning model was accuracy of 96. 1 %, sensitivity of 95.2 %, specificity of 96.9 %, F-value of 0.961, and AUC of 0.99. The cases who were correct for the diagnosis showed generally correct diagnostic basis using Grad-CAM.
[Conclusions]
The CAD system using deep learning model which we developed was able to diagnose hip fracture in the plane X-ray with the high accuracy, and it was possible to present the decision reason.

## 1. Introduction

In Japan, as many as 13 million elderly people have osteoporosis [31,32]. Fragility fractures such as hip fractures and spinal fractures associated with osteoporosis are also increasing, with 200,000 patients suffering from hip fractures annually in particular [23]. Patients with hip fracture are required to be admitted to the hospital as soon as possible because their walking ability and level of daily living are greatly reduced and their vital prognosis is greatly affected [4,9].

Many patients with hip fracture visit the emergency department because they have difficulty walking due to pain. In emergency department, clinicians are exposed to excessive time and mental stress, which can cause fatigue and misdiagnosis [2,11]. In previous studies, the misdiagnosis rate at the initial diagnosis for hip fractures was said to be 7 - 14% [10] [5]. This tendency is particularly pronounced in emergency department where patients with hip fractures present and initial treatment is provided by residents [19]. Delay in diagnosis and treatment worsens prognosis [27], and misdiagnosis can also lead to medical litigation [2]. To prevent misdiagnosis, radionuclide bone scans, computed tomography (CT), and magnetic resonance imaging (MRI) as well as plain X-rays are recommended as additional diagnostic imaging when diagnosis is difficult [10] [24], but they are not ones that can be done urgently in all institutions.

In recent years, deep learning, a method of machine learning using multi-layered neural networks, has emerged and improved the accuracy of computer image recognition [12]. The research which applied this is widely studied in the medical field, and many research results have already been published [17]. The research also advances in the field of orthopedics, and the diagnosis of fracture by artificial intelligence (AI) using the deep learning approach is reported for various parts including ankle joint and wrist joint first reported by Olczak et al in 2017 [15,22]. In addition, a previous study reported that clinicians can significantly improve the diagnosis rate for wrist fractures with computer-aided diagnosis (CAD) system using AI [20]. However, these are the results of a single institution and relatively small size of dataset that developed these systems.

Therefore, in order to carry out the research which assumed clinical use, we created a large data set of plane X-rays of the hip fractures imaged by various protocols collected in the multi-institutional collaboration. Then, the training of the deep learning model was carried out using the data set, and the CAD system which could offer the judgment reason was developed.

## 2. Materials and Methods

### 2.1 Patient Registration

This study was conducted with the approval of the ethics committee of each hospital (Gamagori City Hospital: approval No. 368-1, Tsushima City Hospital: approval No. 2019-3, Nagoya Daini Red Cross Hospital: approval No. 1360). This multicenter collaborative development study collected medical images from 3 hospitals (Gamagori City Hospital, Tsushima City Hospital, Nagoya Daini Red Cross Hospital) in Aichi Prefecture, Japan. The Nagoya Daini Red Cross Hospital is an emergency medical institution in a city with a population of 2.3 million. On the other hand, Gamagori City Hospital and Tsushima City Hospital are emergency medical institutions in local cities with populations of 5 - 70,000, respectively. The background of the three institutions is shown in Table 1.

Table 1. Facility information of participating medical institutions

| | Gamagori City Hospital | Tsushima City Hospital | Nagoya Daini Red Cross Hospital | Overall | P-value |
|---|---|---|---|---|---|
| Medical sphere (people) | 140 thousand | 300 thousand | 570 thousand | 1,010 thousand | <0.01 |
| Number of ambulances per year in 2019 | 3,351 | 4,380 | 12,726 | 20,457 | <0.01 |
| Number of emergency patients in 2019 | 14,131 | 13,724 | 37,713 | 65,568 | <0.01 |
| Number of residents in 2019 | 7 | 11 | 47 | 65 | <0.01 |
| X-ray generator | MODEL TF-6TL-6 (TOSHIBA, Japan) | UD150L-40 (SHIMADZU, Japan) | DHF-153HII (HITACHI, Japan) | | |
| Image file format | .jpeg | .dcm | .jpeg | | |
| Image size | 4892×4020 pixel | 3451×2836 pixel | 2039×1380 pixel | | |

We included 5295 cases of neck fractures or trochanteric fractures that were diagnosed and treated by orthopedic surgeons using plane X-rays or CT or MRI in outpatient visits to each institution between April 2009 and March 2019. Among these 5295 cases, 391 cases suffered hip fracture on one side during the study period and subsequently suffered a hip fracture on the opposite side. We also used X-rays that included pathologic fractures of the proximal femur due to tumors (12 cases), cases with osteoarthritis of the hip (K/L Grade III or higher: 84 cases) [13], images that included hip implants on the opposite side (452 cases), images that included spine implants (46 cases), and images that included obsolete or complicated pelvic fractures at the time of injury (93 cases) to approximate the actual clinical settings. Images in which both hips were not within the imaging range (14 cases), cases of femoral shaft fracture (7 cases), and cases of periprosthetic fracture (32 cases) were excluded. Finally, 5242 plane frontal hip X-rays taken at the time of injury in 4851 cases (Sex: 1193 males, 3658 females, mean age at injury: 81.1 years old (95%CI: 69.6, 92.6)) were used for the study (Figure 1).

2.2 Evaluation of the fracture type

Two orthopedic surgeons (YS, TA) assessed the presence and the type of fracture. The Kappa statistic of inter-rater agreement for the presence or absence of these fractures was 0.91. If the results differed, the two were consulted to determine the presence of a fracture. To classify the fracture type, we used the Garden classification(Garden classification stage I,II,III,IV : G/S I-IV) for neck fractures [3] and the old AO/OTA classification for trochanteric fractures (AO/OTA 31-A1, A2, A3) [14], considering the inter-rater agreement in previous studies. Fractures of the greater trochanter of the femur in which the fracture line did not reach the medial-cortex on MRI were classified as isolated great trochanter fracture [21]. 5024 cases (95.8 %) were diagnosed from hip frontal plain X-rays, 97 cases (1.9 %) from lateral X-rays, and 121 cases (2.3 %) from CT or MRI. Table 2 shows the classification of fracture types. There was a significant difference in age and fracture type in three institutions.

2.3 Image capturing environment and image data extraction

Both hips were photographed using a supine imaging table with both hips rotated inward. The X-ray generator was positioned directly above the center of both hips, and the imaging plate cassette was placed directly under the buttocks, and the center line was perpendicularly incident on the cassette in the anteroposterior direction toward the center of both hips. The imaging conditions were 70 kV tube voltage, 200 mA tube current, 0.4 s imaging time, and 100 cm source image receptor distance. The X-ray generator and CR image processor are shown in Table 1. Digital Imaging and Communications in Medicine (DICOM) image display (Toshiba Medical Systems Corporation, Tokyo) was used as the image reference software, and anonymized image data were extracted from the DICOM server. The image file format and the size of the original image at each institution for data extraction are shown in Table 1.

Table 2. Patient background and fracture type classification

| | | Gamagori City Hospital | Tsushima City Hospital | Nagoya Daini Red Cross Hospital | Overall | P-value |
|---|---|---|---|---|---|---|
| Mean age at time of injury (95%CI) | | 81.8 (70.4,93.2) | 81.4 (70.9,91.9) | 80.1 (67.6,92.6) | 81.1 (69.6,92.6) | <0.01 |
| Sex | Male | 340 | 287 | 566 | 1193 | 0.13 |
| | female | 1,156 | 829 | 1,673 | 3658 | 0.13 |
| Fracture type | Garden I / II | 275 | 191 | 528 | 994 | <0.01 |
| | Garden III / IV | 450 | 324 | 897 | 1671 | <0.01 |
| | AO31-A1 | 489 | 383 | 509 | 1381 | <0.01 |
| | AO31-A2 | 253 | 185 | 322 | 760 | 0.07 |
| | AO31-A3 | 54 | 48 | 76 | 178 | 0.41 |
| | Greater trochanteric fracture | 96 | 74 | 88 | 258 | <0.01 |
| Number of X-rays | | 1617 | 1205 | 2420 | 5242 | |
| Complications | pathologic fractures of the proximal femur due to tumors | 3 | 2 | 7 | 12 | 0.80 |
| | cases with osteoarthritis of the hip | 26 | 15 | 43 | 84 | 0.50 |
| | Cases with hip implants on the opposite side | 132 | 125 | 195 | 452 | 0.05 |
| | Cases with spine implants | 7 | 4 | 35 | 46 | <0.01 |
| | obsolete or complicated pelvic fractures | 23 | 17 | 53 | 93 | 0.12 |

※Age at injury and fracture type were evaluated for each image when there were multiple images in bilateral cases and time series.

Figure 1. Patient flow

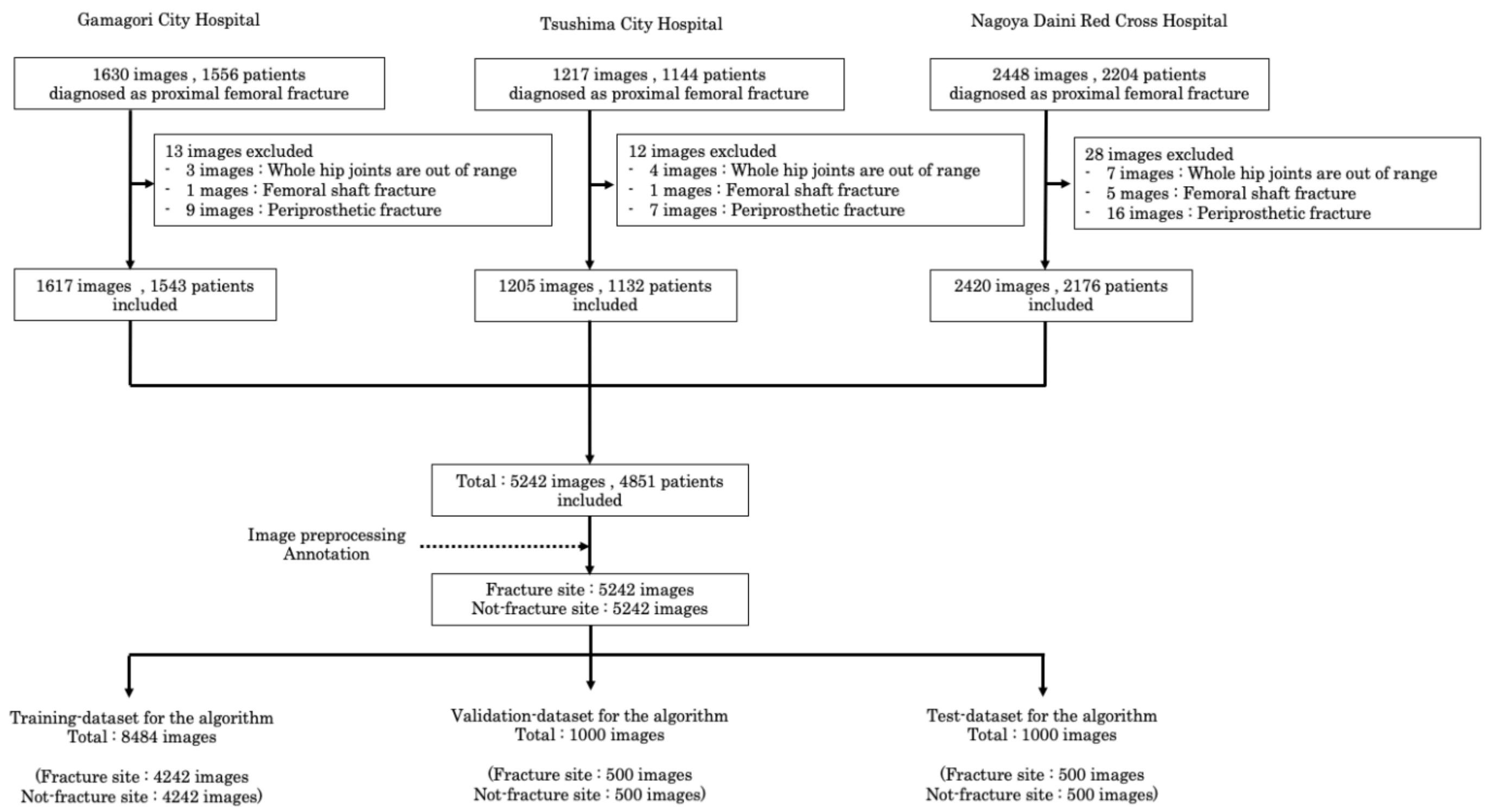

2.4 Image Preprocessing and Development of the algorithm

We used Intel Core i7 8700 K, Ubuntu 18.04, and Python 3.7 to perform image processing on the target image data. Images extracted from the DICOM server were converted into 3 channel * 8 bit JPEG images, and both were resized to 380 * 380 pixels. The data is not compressed. All of the resized images were checked by 2 orthopedic surgeons (YS, TA) and given a rectangle that included the entire fracture site. In order to extract images with a larger area, a vertical dividing line was placed at a position with a 50-pixel margin for the rectangle, and the images without the rectangle were adopted as the non-fractured site data, and 5242 images without the fracture side were generated. The image of the side containing the rectangle with the same size as the non-fractured site data was adopted as the fracture side data, and 5242 images containing the fracture side were generated. In total, 10484 images were prepared for machine learning (Supplementary Figure 1).

We randomly divided the data set into three ones: a training dataset (4242 non-fracture side and 4242 fracture side, for a total of 8484 images), a validation dataset (500 non-fracture side and 500 fracture side, for a total of 1000 images), and a test dataset (500 non-fracture side and 500 fracture side, for a total of 1000 images) (Figure 1).

We used Intel Core i7 8700 K, 32 GB memory, and Ubuntu 18.04 as learning environments. Python 3.7 was used to train an algorithm for the analysis, and deep learning libraries Pytorch 1.3 and Fast.ai 1.0 were used. Nvidia's RTX 2070 GPU was used for learning and reasoning of deep learning. In order to perform transfer learning [26], We also used EfficientNet-B4 [28] which was pre-trained ImageNet model [26] (Supplementary Figure 2). A deep convolutional neural network (DCNN) approach was used for the learning. The model was trained as two-class classification, with images with fractures as positive and images without fractures as negative. The training dataset and the validation dataset were used for the training.

The initial learning coefficients were increased from 0 to WarmStartup, then to about 10e-3 to 1cycle, and then to Decay. The learning time is about 10 minutes per epoch, and the overall time is about one hour in six epochs. We used Adam as an optimizer. The batch size was set at 40, and the validation loss curve was confirmed in about 1200 batches, and it was judged that the plateau in performance was reached. Annealing of the LR was planned and the learning rate decay was performed in one cycle. (Supplementary Figure 3).

In addition to the use of Dropout (p=0.4) in EfficientNet-B4, which is included in the model adopted as a countermeasure against overlearning, we performed random mirroring on the vertical axis as a data augmentation and light and dark changes randomly during learning. Early stopping is not used because of LR decay.

2.6 Considerations

2.6.1 Accuracy evaluation of the learning model

The diagnostic accuracy of the trained training model was evaluated for the test image dataset. Evaluation items are accuracy, sensitivity, specificity, and F-value. We also calculated the receiver operating characteristic (ROC) curve and measured the area under the curve (AUC).

2.6.2 Visualization of the basis for diagnosis of CAD system

The gradient-weighted class activation mapping (Grad-CAM) [25] was adopted to conceptualize the basis for the learning model's diagnosis of a fracture.

The show-heatmap-function of Fast.AI ( http://www.fast.ai ) was run on the training model to obtain the heatmap. To investigate the practicality, we calculated the calculation time for the whole process of the inference and the generation of heat map for one image of test data. The calculation method is the average time per image of test data divided by the calculation time, which was deduced from 1000 images of test data. An orthopedic surgeon (YS) evaluated a heat map accuracy whether the fracture site on the original image was included in its activation area or not. The evaluation method was based on the previous research [6]. The intra-rater agreement in the evaluation is determined over a two week period, with a Kappa statistic of 1.0.

2.7 Statistical analysis and software

The statistical software EZR was used for the statistics [16]. The Mann-Whitney U test was used for normality and non-parametric in the Shapiro Wilk test, and the Fisher's exact test was used to analyze the categories. Significance was set at $p<0.05$. Scikit-Learn ( https://scikit-learn.org/ ) was used to analyze the accuracy of the training model. A 95%CI was calculated for each value.

## 3. Results

3.1 Accuracy evaluation of the learning model

The learning accuracy of the learning model was accuracy of 96.1 % (95%CI: 94.9, 97.3), sensitivity of 95.2 % (95%CI: 93.9, 96.5), specificity of 96.9 % (95%CI: 95.8, 98.0), and F-value of 0.961 (95%CI: 0.950, 0.972). The ROC curve was as shown in the figure, and the AUC of 0.99 (95%CI: 0.98, 1.00) (Figure 2).

Figure 2. ROC Curves

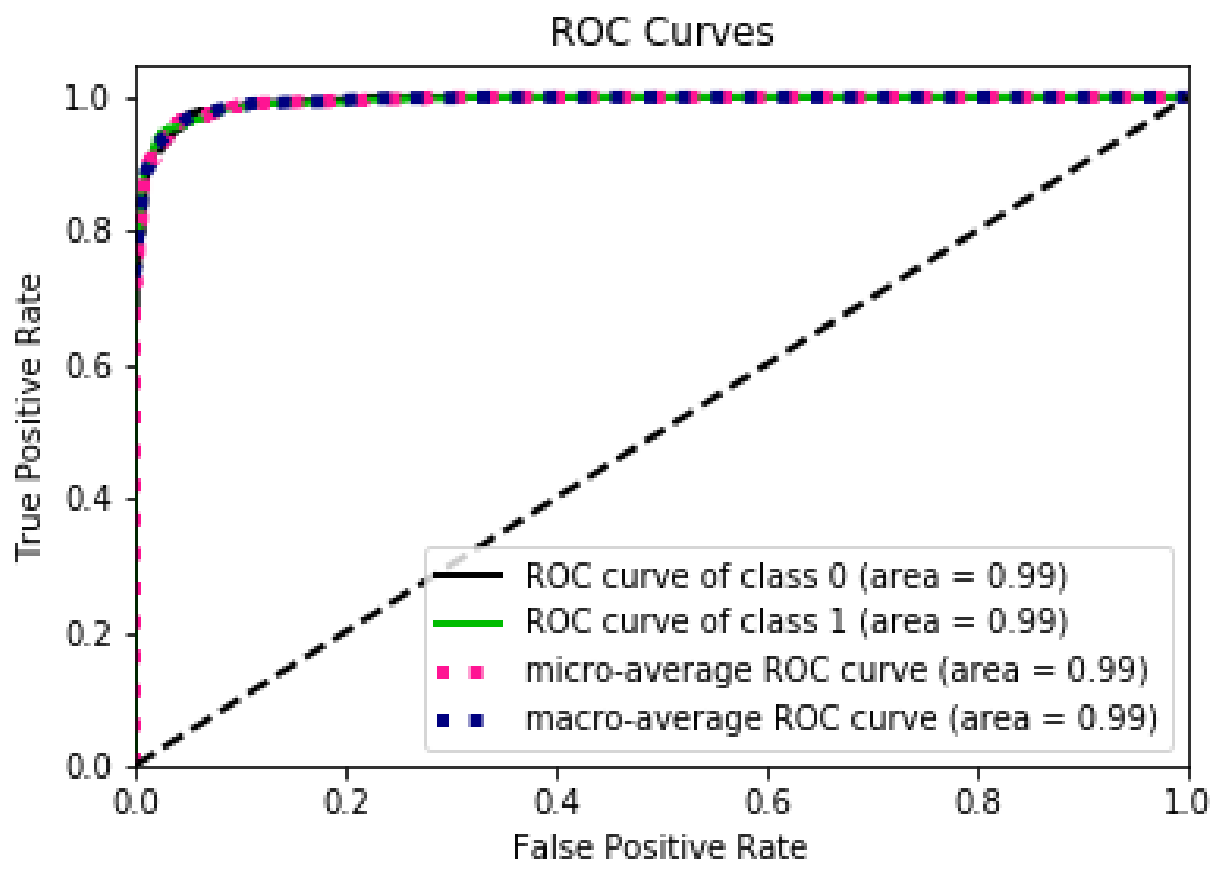

This is the ROC curve for the EfficientNet-B4 model, with an AUC of 0.992. This is the ROC curve for the EfficientNet-B4 model, with an AUC of 0.992.
Class 0 indicates cases without fracture, and Class 1 indicates cases with fracture. Each ROC curve was calculated. Micro-average ROC sums contributions bby class, while macro-average ROC shows the average of results for all classes (AUC = 0.992).

Figure 3. The images that the CAD system misdiagnosed

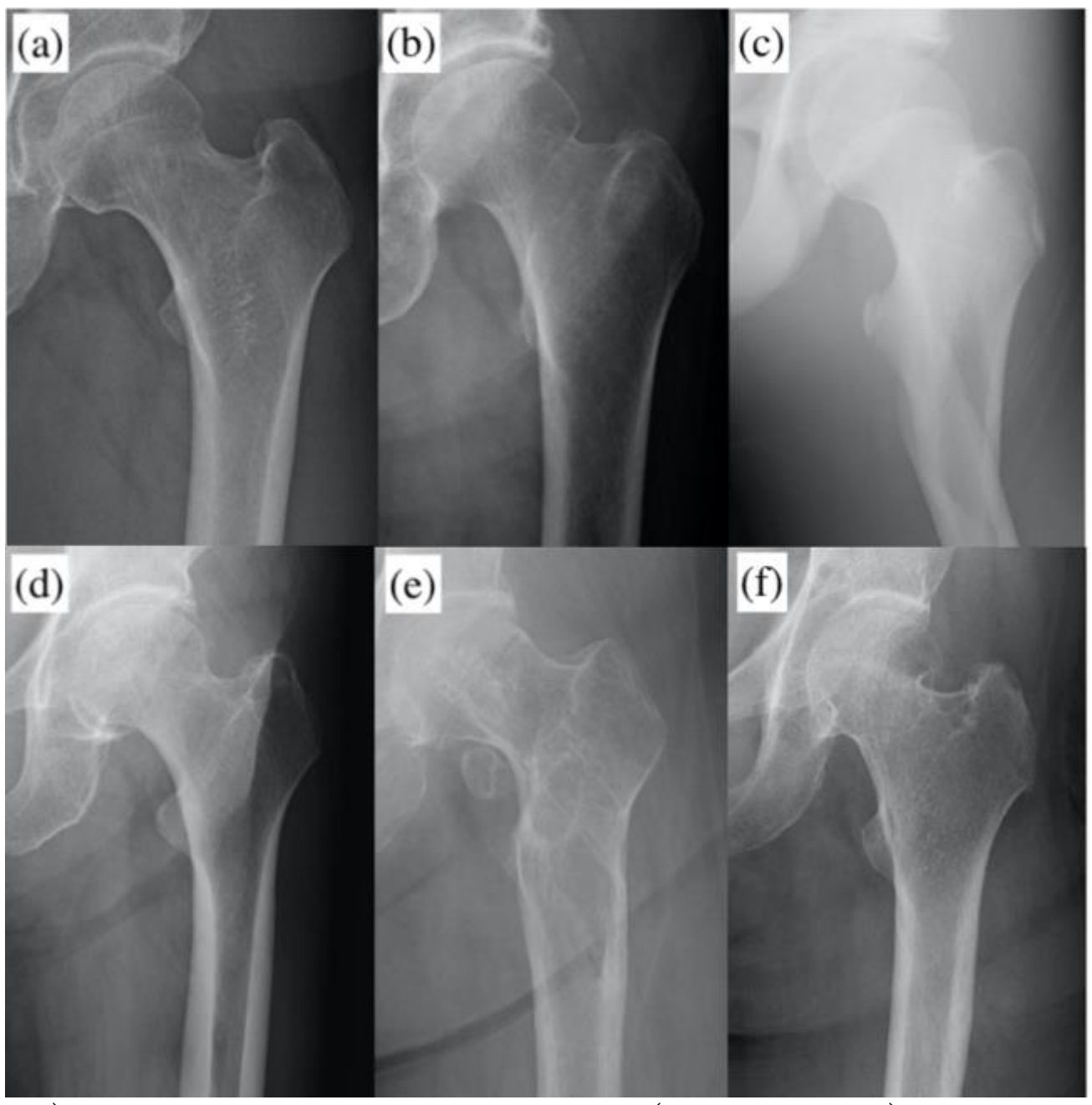

a-c) CAD system incorrectly diagnosed (false negative)
a) Cases that even we orthopedic surgeons are at a loss to decide
b) Cases in which a non-orthopaedic surgeon could be wrong
c) Cases that even non-orthopaedic surgeons are not confused by the diagnosis
d-f) CAD system incorrectly diagnosed (false positive)
d) Normal image.
e) A case after implant removal
f) A case of deformity healing after conservative treatment

On the other hand, the CAD system misdiagnosed 39 images in total of 1000 images. A total of 24 images with fractures were diagnosed as "without fracture" (false negative). The images consisted of 21 fracture types with relatively slight displacement (neck fracture (G/S 1,2); 8 fractures, trochanteric fracture (AO31-A1); 4 fractures, fractures of the greater trochanter of the femur; 9 fractures), and 3 fracture types with relatively large displacement (neck fracture (G/S 3,4); 1 fracture, trochanteric fracture (AO31-A2,3); 2 fractures). A total of 15 images without fracture were diagnosed as "with fracture" (false positive). There was a case of hip fracture with deformity after conservative treatment, a case after nail removal, and 13 cases with normal image (Figure 3).

3.2 Visualization of the basis for diagnosis of CAD system

In the accuracy verification using Grad-CAM for images with fractures, all the images that were correctly diagnosed depicted feature points consistent with the truly fracture site (Figure 4). The average inference time per 1 image including Grad-CAM was 1.17 seconds.

Figure 4. Visualization of fracture detection area using Grad-CAM.

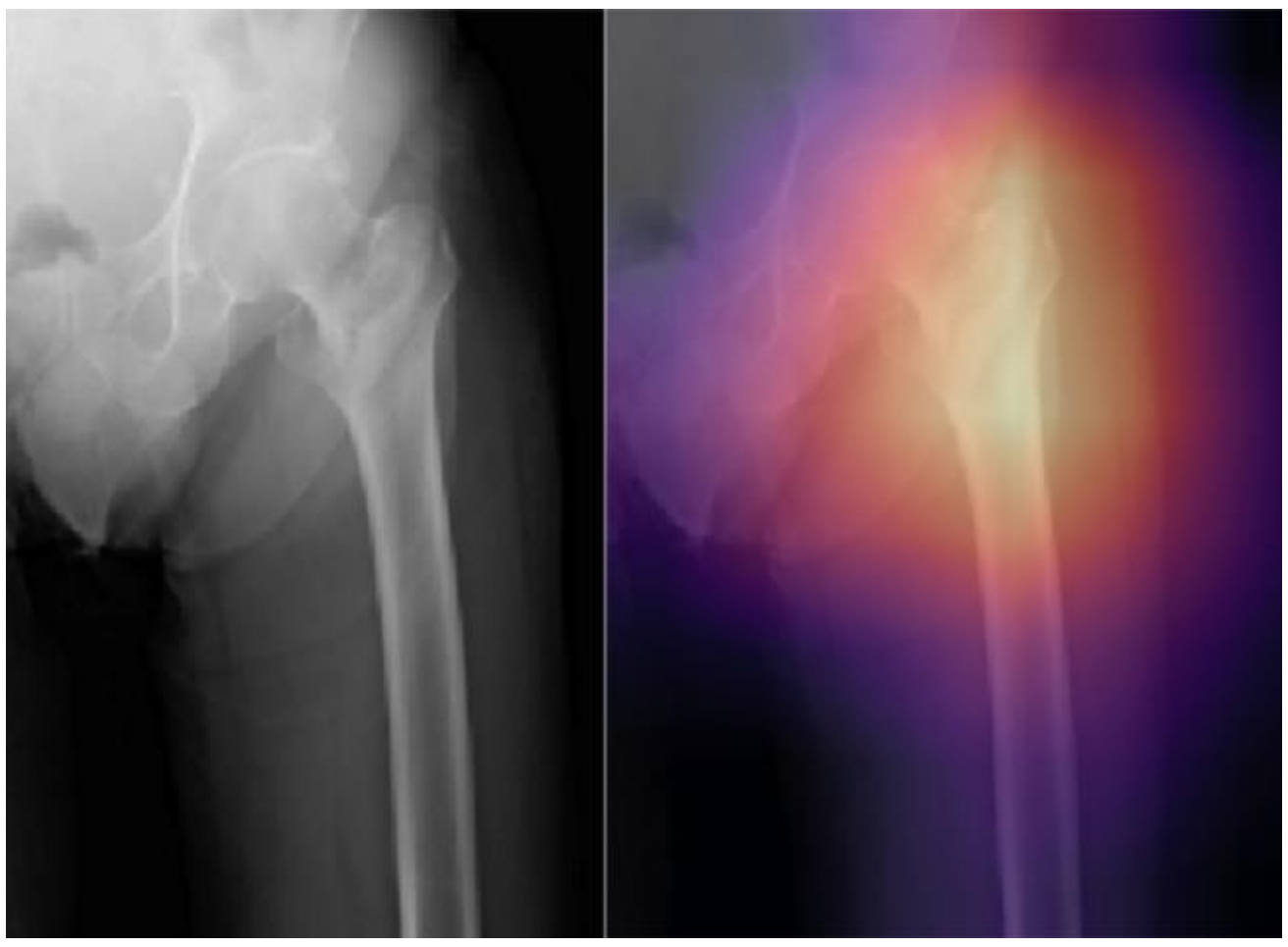

The CAD system in this study was able to extract the feature points of the images that diagnosed the fracture appropriately (left: original image, right: Grad-CAM image showing the basis of the fracture site as a heat map). From yellow to red, the diagnostic basis of the CAD system is strongly evident.

## 4. Discussion

In this study, we show that an AI-based CAD system can achieve high diagnostic accuracy for hip fractures. Then, we were able to generate a heat map for the fracture site to provide a diagnostic basis.

The diagnostic accuracy of AI-based CAD system in this study was as good as that reported by others, and it was possible to show the highest AUC among previous studies for hip fractures [1,6,29] (Table 4). The first strength of this study is that a large amount of learning data was secured. Although large data sets are considered the key to success in machine learning [8], the number of data that can be collected is limited in the case of single-center studies, and training on small data sets may result in overlearning [30]. The majority of the published studies on AI so far have been conducted by a single institution, and only 6% of them reported assessing the utility of data extracted from different environments [18]. The advantages of multicenter research are 1) the ability to collect a large amount of data, and 2) the ability to conduct research from data obtained from different conditions and environments to conduct reasonable medical research which assuming the use in actual clinical settings [18]. In this study, almost all images of hip fractures from multiple institutions were used, and about 10,000 images of machine learning data were generated from about 5,000 cases. This allowed us to prepare a dataset that included a wide variation of hip conditions for each patient, including individual differences in pelvis and femur. By preparing sufficient training data, the training model was able to correctly diagnose patient-specific information other than fractures as "negative", such as implants on the opposite side of femur or spine. And, in this study, high accuracy was achieved even with different radiographic equipment and image file formats at multiple institutions. We believe that this study will contribute to the practicality of the CAD system, considering that the CAD system will be used in actual clinical settings with versatility in the future.

On the other hand, there were 3.9 % of images ( 39 out of 1,000 test data) in which the AI failed to diagnose correctly, and 24 of them were misdiagnosed as "no fracture" despite the presence of a fracture. In the diagnostic imaging test in this study, the sensitivity for fracture diagnosis was comparable between the learning model and orthopedic surgery fellows (95.8 %: 95.5 %). This suggests that AI does not have image diagnosis ability beyond those of orthopedic surgery fellows.

Second, our system was able to provide a diagnostic basis by generating a heat map for the fracture site. And, the results were consistent with the fracture site indicated by the orthopedic surgeon. It was necessary to solve the "black box problem" unique to AI when assuming its practical application. Deep learning used in image recognition realizes the classification for the data which cannot express the feature quantity explicitly originally, and the reason of the judgment is uncertain, and human cannot understand and interpret it. This is called the black box problem [7]. In this study, we used Grad-CAM to visualize class-discriminative regions on the x-rays when the learning model judged. When evaluated against this, the class identification region was consistent with the fracture site. This could provide a reason to actually diagnose a hip fracture, although it is still unclear which of the image information the learning model based its decision on, such as the fracture line, bone marrow edema, and soft tissue contrast. This may increase the reliability of the AI proposal when CAD is used as a diagnostic aid.

There are several limitations of this study.

First, the present dataset includes cases of pathological fractures caused by metastatic bone tumors, but does not include cases that do not have fractures but have osteomyelitis. It is desirable to consult a specialist as soon as possible in such cases, but the CAD system developed in this study may not be able to point this out.

Table 4. Litrature review

| | Year | Institute | Number of patients | Number of hip images | Images including implant on hip or spine | Model of transfer learning (parameter) | Accuracy | Sensitivity | Specificity | AUC | Grad-CAM | Clinician test (AI-aided test) |
|---|---|---|---|---|---|---|---|---|---|---|---|---|
| Adams et al [1] | 2018 | 1 | N/A | 805 | excluded | GoogLeNet (6.8M) | 90.6% | N/A | N/A | 0.98 | no | no |
| Urakawa et al [29] | 2018 | 1 | 1773 | 3346 | excluded | VGG_16 (138M) | 95.5% | 93.9% | 97.4% | 0.969 | no | no |
| Cheng et al [6] | 2019 | 1 | 3605 | 3605 | included | DenseNet-121 (8M) | 91.0% | 98.0% | 84.0% | 0.98 | yes | no |
| Current Study | 2020 | 3 | 4851 | 10484 | included | EfficientNet-B4 (19M) | 96.1% | 95.2% | 96.9% | 0.99 | yes | yes |

Second, the image needs to be divided as a preprocessor. A CAD system that can diagnose hip fractures without preprocessing from X-rays of both hips should be redeveloped using the learning model obtained in this study.

Third, the external validity is not guaranteed. This study is a retrospective evaluation performed through a web interface similar to PACS used by clinicians for medical imaging. It is possible that the incidence of "with fracture" images in clinical practice differs from the frequency of diagnosis in actual clinical settings. Prospective studies using the PACS system in an actual clinical setting should be carried out in the future.

## 5. Conclusion

Our CAD system using AI for hip fracture has become an imaging tool with high diagnostic accuracy as well as providing a diagnostic basis.

## 6. Conflict of interest statement

The authors, RG and AK, are employees of Search Space CO.Ltd., a startup company, the eventual products and services of which will be related to the subject matter of the article

The all authors do not own shares in the above companies. SH, the last author, represents the AI research division in the nonprofit organization (NPO) Nagoya Orthopedic Regional Healthcare Support Center, ( https://www.fracture-ai.org/ ). NPO Nagoya Orthopedic Regional Healthcare Support Center, AI Research Division is a research division established for multi-center collaborative research. With the exception of two Search Space CO.Ltd employees and one NPO employee, the other authors received no compensation from these organizations.

## 7. Ethical review committee statement

This study was approved by each institutional review board and all experiments were performed in accordance with the ethical standards laid down in the amended declaration of Helsinki. This study was conducted with the approval of the ethics committee of each hospital. (Gamagori City Hospital: approval No. 368-1, Tsushima City Hospital: approval No. 2019-3, Nagoya Daini Red Cross Hospital: approval No. 1360)

## 8. A statement of the location where the work was performs

Image labeling and image preprocessing were performed in the hospitals of Gamagori City Hospital and Tsushima City Hospital, and machine learning was performed internally at Search Space CO,Ltd. A questionnaire service on the Internet ( Google Form, https://docs.google.com/forms/ ) was used to conduct the diagnostic imaging test among clinicians.

Supplemental figure 1. Image preprocessing

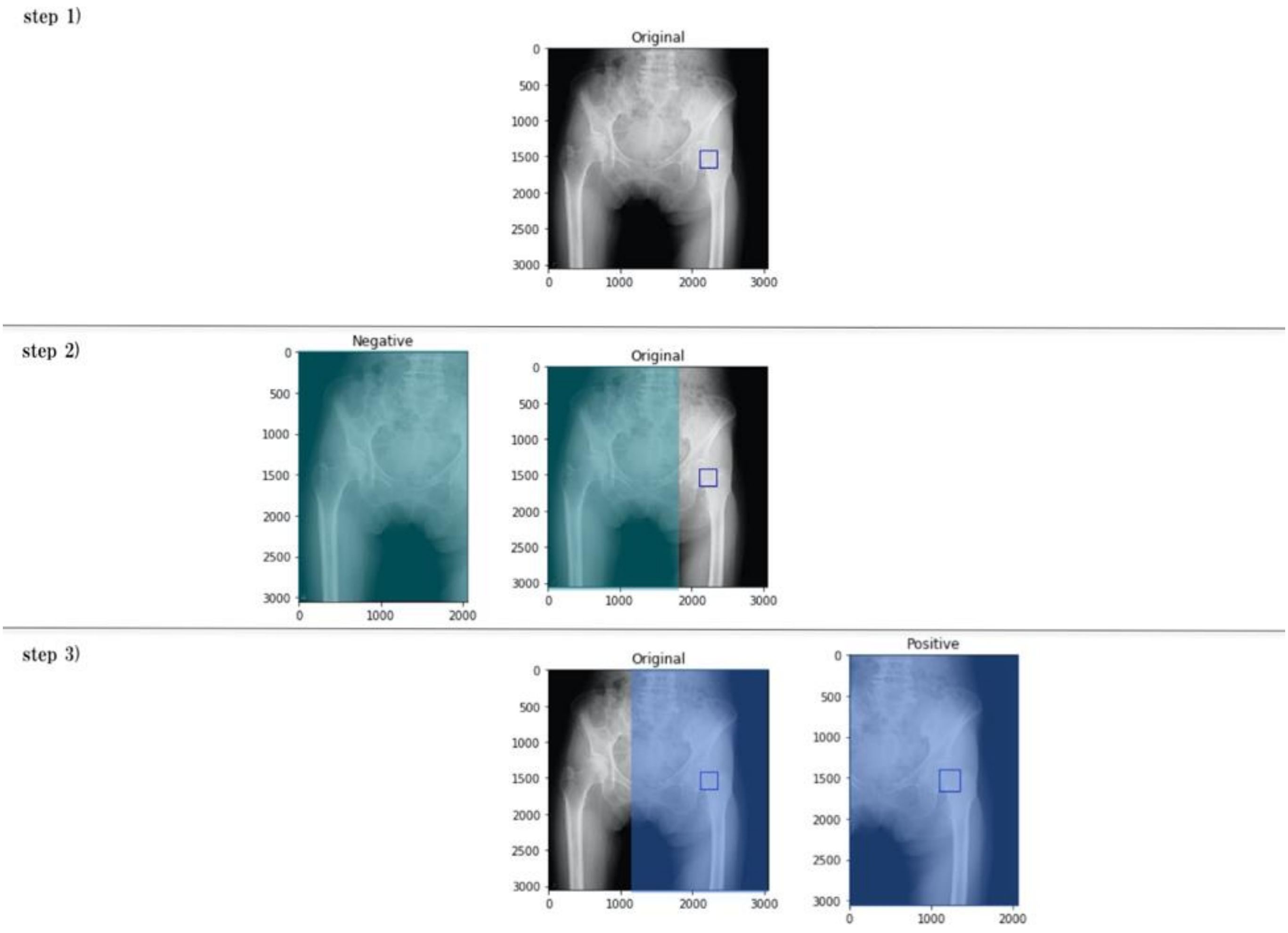

Step 1) For all 5242 plane X-rays, a rectangle of the shape including the fracture area was assigned by orthopedic surgeons.
Step 2) A margin of 50 pixels from the rectangle was set up and a dividing line was inserted, and the image without the rectangle was used as the non-fractured image.
Step 3) The image of the fractured side is the same size as the non-fractured side and includes a rectangle.

Supplemental figure 2. Configuration diagram of EfficientNet-B4 model

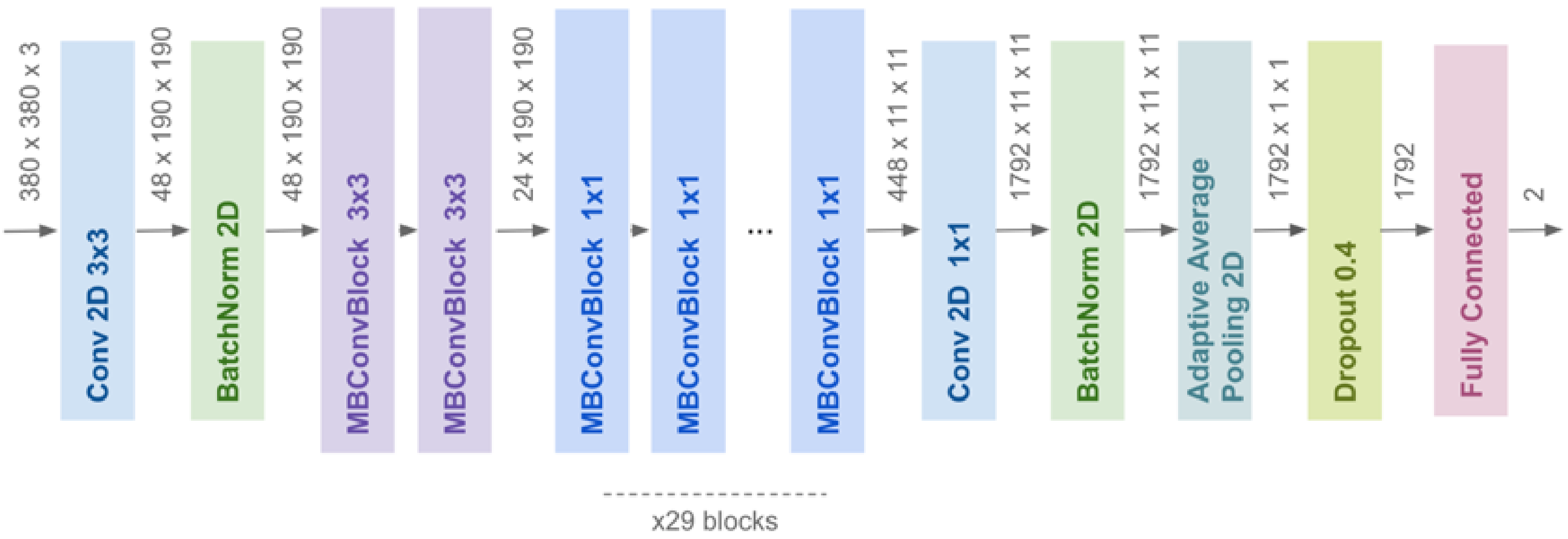

The EfficientNet-B4 model used in this paper combines the depth, breadth, and input resolution of neural networks with the best efficiency to performance ratio in existing studies. It was adopted because the number of parameters is small compared to other learning models, the model is simple, and it is suitable for transfer learning.

Supplemental figure 3. Machine learning process

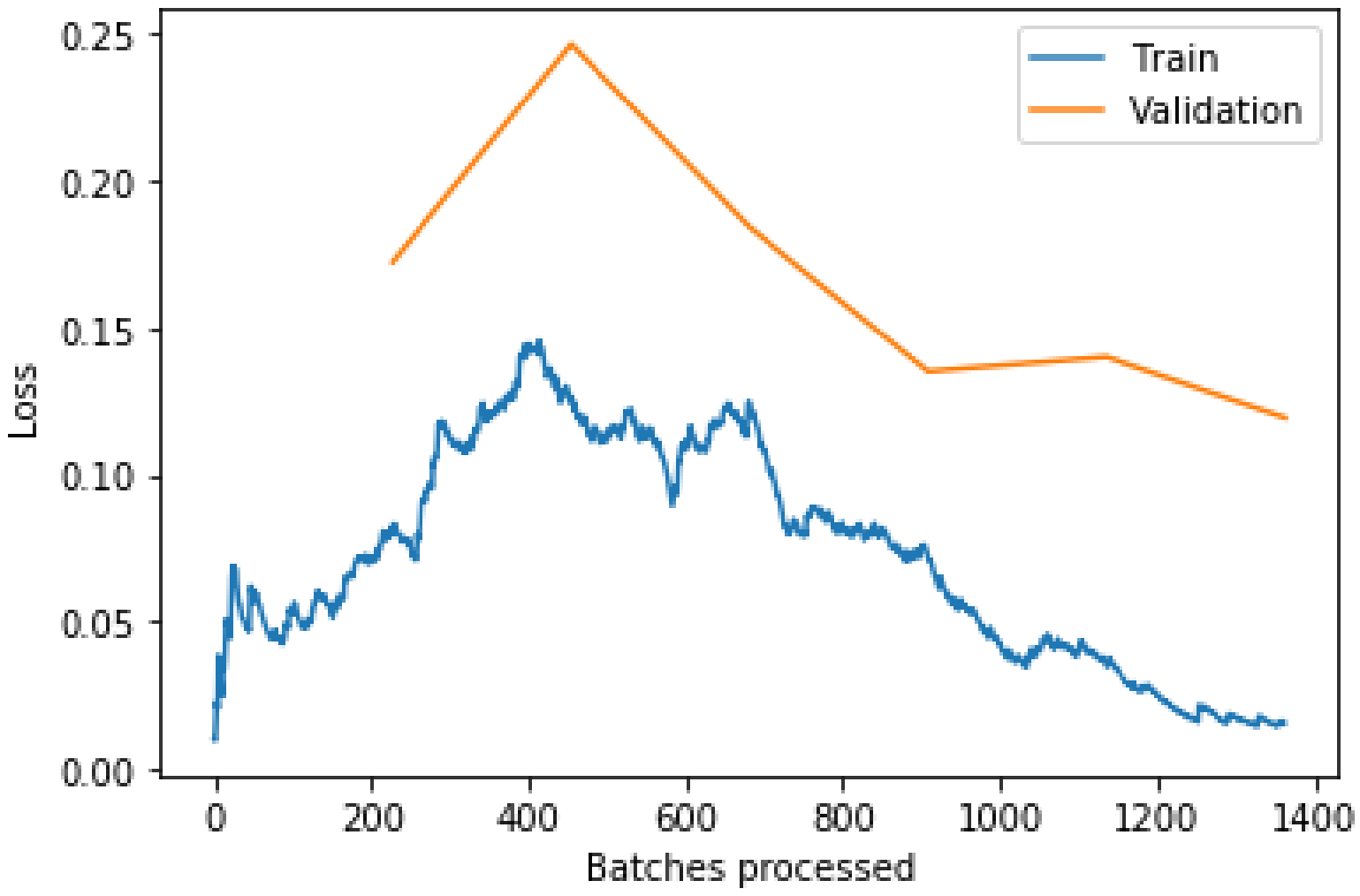

Learning processes in training and validation image datasets. 1 batch = 40 plane X-ray images, 8484 training image datasets were trained six times, and the errors confirmed by the validation image datasets became smaller and smaller as the training progressed. The training time was about one hour.